\documentclass{elsart}

\usepackage{graphicx}
\usepackage{amssymb}
\usepackage{epsfig}
\usepackage{subfigure}

\usepackage[english]{babel}
\usepackage{latexsym}
\usepackage{amsfonts}
\usepackage{amsmath}
\usepackage{multirow}
\usepackage{float}
\usepackage{longtable}

\def\be{\begin{equation}}
\def\ee{\end{equation}}

\begin{document}
\begin{frontmatter}

\title{Statistical hadronization of heavy flavor quarks in elementary collisions:
  successes and failures}

\author[gsi]{A.~Andronic},
\author[hei]{F.~Beutler}
\author[gsi,emmi,tud,fias]{P.~Braun-Munzinger},
\author[wro,tud]{K.~Redlich},
\author[hei]{J.~Stachel}

\address[gsi]{GSI Helmholtzzentrum f\"ur Schwerionenforschung,
D-64291 Darmstadt, Germany}
\address[hei]{Physikalisches Institut der Universit\"at Heidelberg,
D-69120 Heidelberg, Germany}
\address[emmi]{ExtreMe Matter Institute EMMI, GSI, D-64291 Darmstadt, Germany}
\address[tud]{Technical University Darmstadt, D-64289 Darmstadt, Germany}
\address[fias]{Frankfurt Institute for Advanced Studies, J.W. Goethe University,
D-60438 Frankfurt, Germany}
\address[wro]{Institute of Theoretical Physics, University of Wroc\l aw,
PL-50204 Wroc\l aw, Poland}

\begin{abstract}
  We analyze recently compiled data on the production of open heavy flavor
  hadrons and quarkonia in e$^+$e$^-$ as well as pp and p-nucleus collisions in
  terms of the statistical hadronization model. Within this approach the
  production of open heavy flavor hadrons is well described with parameters
  deduced from a thermal analysis of light flavor hadron production. In
  contrast, quarkonium production in such collisions cannot be described in
  this framework. We point out the relevance of this finding for our
  understanding of quarkonium production in ultra-relativistic nucleus-nucleus
  collisions.
\end{abstract}

\end{frontmatter}

\section{Introduction}

One of the major goals of ultrarelativistic nuclear collision studies is to
obtain information on the QCD phase diagram \cite{pbm_wambach}.  A promising
approach is the investigation of hadron production.  Hadron yields measured in
central heavy ion collisions can be described very well (see ref. \cite{aa08}
and references therein) within a hadro-chemical equilibrium model.  
In our approach \cite{aa08}, the only parameters are the chemical freeze-out
temperature $T$ and the baryo-chemical potential $\mu_b$ (and the fireball
volume $V$, in case yields rather than ratios of yields are fitted), 
characteristic for a given energy; for a review see \cite{review}.

Focused on production of hadrons carrying light ($u$, $d$, $s$) quarks, 
these investigations led to temperature values which rise rather sharply from 
low energies on towards $\sqrt{s_{NN}}\simeq$10 GeV and reach afterwards
a plateau  near $T$=165 MeV, while the baryochemical potential decreases 
smoothly as a function of energy.
This  limiting temperature  behavior reminds of the Hagedorn
temperature \cite{hagedorn85} and suggests a connection to 
the phase boundary. It was, indeed, argued \cite{wetterich} that the 
quark-hadron phase transition drives the equilibration dynamically, at least 
for SPS energies and above. 

Recently, the results of ref.~\cite{aa08} strenghtened the interpretation 
that the phase boundary is reflected in features of the hadron yields in
nucleus-nucleus collisions. 
Whether the chemical freeze-out curve for $T<$160 MeV traces the QCD phase 
boundary at large values of chemical potential is an open question.
Possibly, the chemical freeze-out in this regime is influenced by exotic 
new phases predicted in \cite{mclerran_pisarski}.

An important  question is whether this  statistical behavior is a unique 
feature of high energy nucleus-nucleus collisions or whether it is also 
encountered in elementary collisions, where finite size effects will obscure
a possible phase transition.
In early analyses (see ref.~\cite{becattini97}), it is indeed argued that 
hadron production in e$^+$e$^-$ and pp is thermal in nature. 
Furthermore, such analyses of hadron multiplicities (for recent results 
see \cite{aa08_ee,becattini08,kraus08} and refs. therein) yield also 
temperature values in the range of 160-170 MeV.

Consequently, alternative interpretations for the apparent statistical
behavior were put forward.
These include conjectures that the thermodynamical state is not reached by
dynamical equilibration among constituents but rather is a generic fingerprint
of hadronization \cite{stock,heinz}, or is a feature of the excited QCD vacuum
\cite{castorina}. In such approaches the relation between hadronization and
the QCD phase transition is not easy to make explicit.
Recent results employing the gauge/string theory duality imply also a thermal
behavior \cite{hatta}, but remain to be further understood.

In our recent analysis of hadron production in e$^{+}$e$^{-}$ collisions
\cite{aa08_ee} we have demonstrated that such a direct connection between
thermal descriptions of hadron production in e$^{+}$e$^{-}$ and relativistic
nucleus-nucleus collisions is much less convincing than thought before. 
We first note that any thermal description of hadron production in 
e$^{+}$e$^{-}$ collisions makes use, in addition to the thermal parameters $T$,
and $V$, of a set of additional, non-thermal parameters such as the number 
of strange, charm, and bottom quark jets as well as an additional strangeness 
(but not charm!) suppression factor $\gamma_s$. 
For details see \cite{aa08_ee,becattini08}. 
Furthermore, we have demonstrated \cite{aa08_ee} that, in spite of the 
additional, non-thermal parameters, the quality of the fit of e$^{+}$e$^{-}$ 
data to statistical model calculations is significantly worse than that 
obtained in descriptions of central nucleus-nucleus collisions.
This conclusion is reached also when the fit to the data is performed 
using the same set of particles as in the nucleus-nucleus case. 
These findings call an overall thermal origin of particle production in 
e$^{+}$e$^{-}$ collisions into question, despite the presence of statistical
features.

In the present paper we explore in some detail the heavy-quark ($c$ and $b$)
sector, i.e. the production of the corresponding open and hidden flavor
hadrons in e$^{+}$e$^{-}$ as well as pp, p$\bar{\mathrm{p}}$ and p-nucleus 
collisions. The emphasis
of our investigations is to establish quantitatively similarities and
differences in the production of such particles in elementary collisions
compared to those observed in the nucleus-nucleus case. 
Since the proposal \cite{satz} that the J/$\psi$ meson production may be 
a 'smoking gun' observable for the diagnosis of the Quark-Gluon Plasma (QGP) 
produced in ultra-relativistic nucleus-nucleus collisions, intense research 
was focused on the topic, both experimentally and theoretically. 
A recent summary is found in \cite{satz_kluberg,rapp_rev}. 

Because the mass of heavy quarks exceeds the transition
temperature of the QCD phase transition by more than a factor of 5, heavy
flavor hadron production cannot be described in a purely thermal approach. 
It was, however, realized in \cite{pbm1} that charmonium and charmed hadron
production can be well described by assuming that all charm quarks are
produced in initial, hard collisions while charmed hadron and charmonium
production takes place at the phase boundary with statistical weights
calculated in a thermal approach. For a recent summary of this statistical
hadronization approach see \cite{pbm_js_lb}. 

In this context it became clear that even complete J/$\psi$ melting in the QGP
via Debye screening as assumed in the original proposal of \cite{satz} could
lead to large J/$\psi$ yields due to production at the phase boundary.
Predictions using the corresponding statistical hadronization model 
(SHM)\footnote{Within this statistical model heavy quarks are not chemically
equilibrated, but otherwise all hadrons are thermalized. The term 'statistical'
is used in this sense in our approach.},
either in its ``minimal'' implementation \cite{pbm1,gor,aa03,aa07} or with
additional assumptions \cite{gra} proved quite successful when compared to
data.  In its application to charm quarks, the statistical model contains as
input the charm production cross section, taken from pQCD calculations
\cite{cac} or from experiment (see ref.~\cite{aa07} and ref. therein).  In
general, statistical production can only take place effectively if the charm
quarks reach thermal (but not necessarily chemical) equilibrium. Combination
of charm quarks which are initially separated by a few fm (corresponding to
about 1 unit in rapidity) into charmonia effectively implies
deconfinement.

To establish the uniqueness of the J/$\psi$ probe for the diagnosis of a QGP in
nucleus-nucleus collisions  it becomes important to understand whether similar 
thermal features as observed in nucleus-nucleus collisions are also 
at work in heavy-flavor hadron production in elementary collisions.

Our paper is consequently organized as follows:
We first briefly describe the model used to analyze particle production in
elementary collisions. Section 3 deals with experimental results
on heavy flavor hadron production in e$^+$e$^-$ collisions and their analysis
in terms of our statistical approach. The more complicated case of pp and
p-nucleus collisions is treated in section 4, with particular emphasis on
hidden charm production. In the final section we summarize our findings and
provide an assessment of charmonium production as a probe for QGP and the QCD
phase boundary.

\section{The model}

For the study of hadron production in e$^+$e$^-$ collisions we employ the 
canonical statistical model described in \cite{aa08_ee,redlich09} (see also 
\cite{becattini08}).
For the present study, we perform calculations for two cases: 
i) a 2-jet initial state which carries the quantum 
numbers of the 5 flavors, with the relative abundance of the five flavors 
in one jet and corresponding antiflavor in the other jet
taken form the measurements at the $Z^0$ resonance quoted in \cite{pdg}.
These relative abundances (17.6\% for $u\bar{u}$ and $c\bar{c}$ and 21.6\% for 
$d\bar{d}$, $s\bar{s}$ and $b\bar{b}$) are thus external input values, unrelated 
with the thermal model.
ii) a purely thermal ansatz , i.e. a 2-jet initial state characterized by 
vanishing quantum numbers in each jet.

For the case of hadron production in elementary hadronic collisions we employ 
the canonical realization of the thermal model \cite{review,gor,aa03,aa07}.
For the description of the relative production cross sections of heavy flavored
hadrons, the energy dependence of the temperature parameter is the only model 
input, which is taken in a parametrized form from the fits of hadron 
abundancies in central nucleus-nucleus collisions \cite{aa08}.
For c.m. energies beyond 10 GeV per nucleon pair in nucleus-nucleus collisions 
a limiting temperature $T_{lim}$=164$\pm$5 MeV is reached. 
Recent fits of hadron yields in pp collisions \cite{kraus08} give very similar
values, independent of anergy. 
The charm production cross section, which is an important model input
parameter for the calculations of absolute yields \cite{aa03,aa07}, cancels out 
for the ratios considered in the present paper.
The influence of the mass spectrum on particle production has been considered
in \cite{aa08}. We note that, for the ratios considered here, such an effect
cancels out in first order and has been neglected.

\section{Results in e$^+$e$^-$ collisions}

\begin{figure}[htb]
\centering\includegraphics[width=.6\textwidth,height=.53\textwidth]{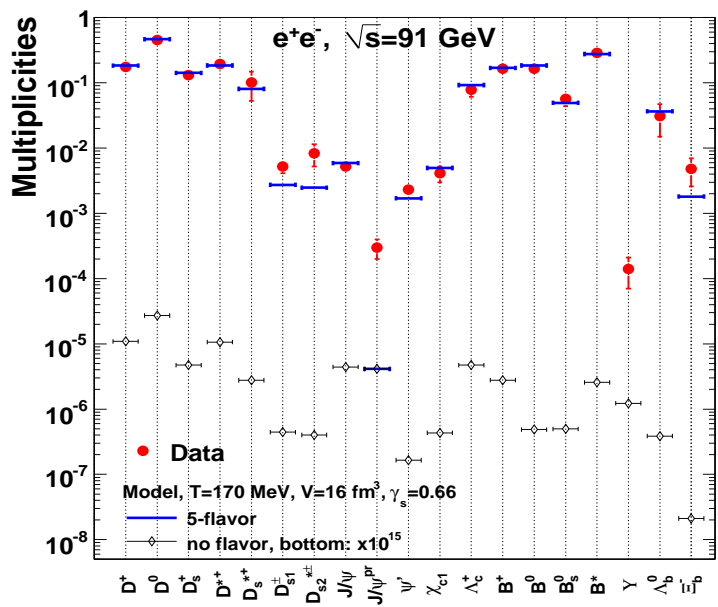}
\caption{Multiplicities of hadrons with charm and bottom quarks in  e$^+$e$^-$ 
collisions compared to the thermal model calculations for two cases: 
i) the 5-flavor jet scheme (thick lines) and ii) no (net) flavor jet scheme 
(thin lines with diamonds). Note, for case ii) the factor 10$^{15}$ used to
scale the model calculations for bottom hadrons to fit in the plotting range.
The data are from the compilation published by the Particle Data Group (PDG) 
\cite{pdg}. The prompt J/$\psi$ measurement J/$\psi^{pr}$ is from the 
L3 experiment \cite{l3_qq}.}
\label{fig1}
\end{figure}

In Fig.~\ref{fig1} we show a comparison of data \cite{pdg} and model prediction
for charmed and bottom hadron yields in e$^+$e$^-$ annihilations at 
$\sqrt{s}$=91 GeV.
For the model we have used the parameter set: $T$=170 MeV, $V$=16 fm$^3$ 
and $\gamma_s$=0.66, which represents the best fit of multiplicities
of hadrons with lighter quarks \cite{aa08_ee}.

We first note that the calculation employing the 5-flavor scheme is in 
very good agreement with the data, as demonstrated by the good $\chi^2$ 
per degree of freedom between the model and the data (excluding the $\Upsilon$ 
and prompt J/$\psi$) of 21.7/16 (34/18 when including all species).
This confirms the conclusion of ref.~\cite{becattini08}.
Despite this overall agreement, the exceptions are significant: 
the $\Upsilon$ meson yield is underpredicted by the model by 17 orders 
of magnitude, while the prompt J/$\psi$ yield \cite{l3_qq} is underpredicted 
by almost 2 orders of magnitude.
Obviously, the production of quarkonia is expected to be strongly suppressed 
in the statistical model.
The disagreement is a consequence of the separate hadronization of the
$c$ and $\bar{c}$ quarks.
The measured prompt J/$\psi$ production in $Z^0$ decays (into hadrons) 
is about 3$\times10^{-4}$ \cite{l3_qq}.
The thermal model predicts a prompt yield for J/$\psi$ of 4.1$\times10^{-6}$
(1.6$\times10^{-7}$ for $\psi'$ and 4.3$\times10^{-7}$ for $\chi_{c1}$),
identically for the two calculation schemes.
The overall measured yields of charmonia are dominated by the feed down 
from bottom hadrons and the model agreement only reflects the agreement seen 
for the open bottom hadrons and their branching ratios to charmonia, 
properly considered in the model.

The calculation employing a purely thermal ansatz underpredicts all
the measurements by many oders of magnitude, while for the light quark sector 
the differences between calculations with a pure thermal model and with 
the 5-flavor quark-antiquark scheme were found to be small \cite{aa08_ee}.
The strangeness suppression factor, which for the present results  
only enters in the calculation of the yields of $D_s$ and $B_s$ mesons, 
appears to have no counterpart in the heavy quark sector.
This reflects the fact that a negligible number of $c$ and $b$ quarks are
formed in the fragmentation process. In this case, the thermal weights 
describe the distribution of the initial quarks into heavy flavor hadrons.
Thermalization is not required in this process.

\section{Results in elementary hadronic collisions}

\begin{figure}[htb]
\centering\includegraphics[width=.56\textwidth]{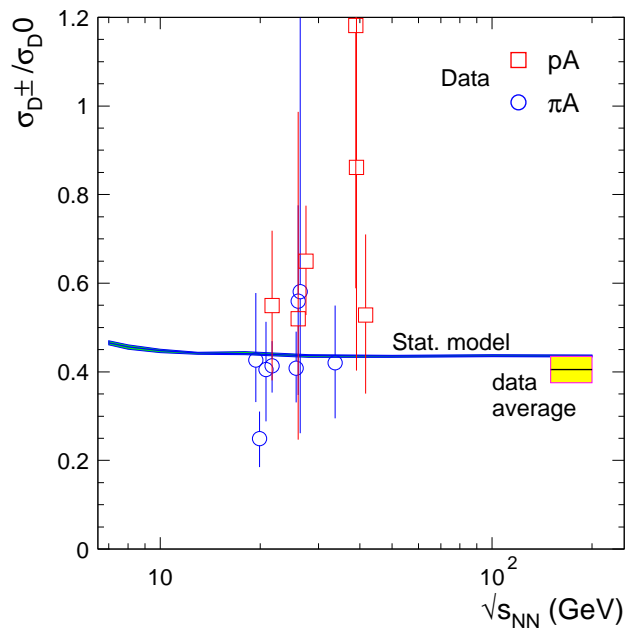}
\caption{Relative production cross section of charged to neutral D mesons.
The data (symbols) in pA and $\pi$A collisions \cite{lou} are compared to
statistical model calculations \cite{aa07} shown by the line
(band corresponding to $\pm$5 MeV errors in $T$). The box is the
average value of all the data points with the corresponding error.}
\label{fig2}
\end{figure}

We now turn to elementary hadronic (pp, p$\bar{\mathrm{p}}$, $\pi$A and pA) 
collisions.
We note an important difference in this case compared to e$^+$e$^-$
collisions, namely that the feeding from bottom hadrons into charmed
hadrons is small at presently available energies due to the much smaller 
bottom production cross section\footnote{In hadronic collisions the relative 
production cross sections between bottom and charm are much smaller than in 
e$^+$e$^-$ collisions. Even at the LHC energies $\sigma_b/\sigma_c\simeq$1/10, 
while in 91 GeV e+e- B(Z$^0\rightarrow b$)/B(Z$^0\rightarrow c$)=0.22/0.17.}. 
In particular, this applies also to charmonia, although at LHC about
20\% of the J/$\psi$ yield is estimated to originate from B mesons decays.

In Fig.~\ref{fig2} we show, as a function of energy, the model comparison to data 
for the relative production cross section of charged to neutral D mesons. 
The data are from the recent compilation of ref.~\cite{lou}.
Considering the relatively large experimental errors, the agreement is good.
This can be judged from the comparison to the average value \cite{pdg} of the 
data points (0.405$\pm$0.030, with a $\chi^2$/d.o.f. of 1.12), shown as well 
in Fig.~\ref{fig2}.
This production ratio is largely determined by feed-down from the 
D$^*$ states (see discussion in \cite{lou}) and the agreement reflects the 
good description within the model of the relative production of the D$^*$ states.

\begin{figure}[htb]
\centering\includegraphics[width=.56\textwidth]{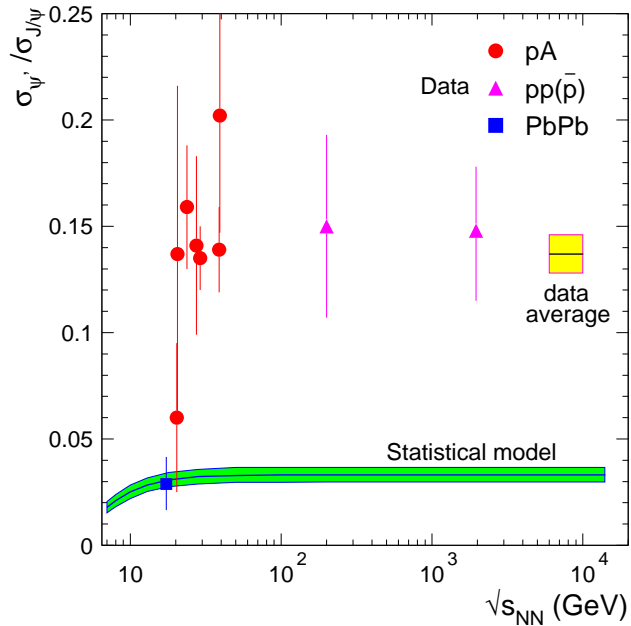}
\caption{Production cross section of $\psi'$ relative to $J/\psi$.
The data for pA collisions are from the compilation by Maltoni et al. 
\cite{maltoni}; the points for elementary collisions are from the PHENIX 
experiment at RHIC \cite{phenix_psip} and from the CDF experiment at Tevatron
\cite{cdf_psip} (see text); the data point for Pb+Pb collisions at the SPS 
energy is from the NA50 experiment \cite{na50}.
The average value of the  pA and pp($\bar{\mathrm{p}}$) measurements with the 
corresponding error (see text) is represented by the shaded box.
The band denotes statistical model calculations \cite{aa07} for the 
temperature parametrization from heavy-ion fits \cite{aa08} ($T_{lim}$=164 MeV) 
with $\pm$5 MeV error.}  
\label{fig3}
\end{figure}

In Fig.~\ref{fig3} we show the model comparison to data for the relative 
production cross section of $\psi'$ and $J/\psi$ charmonia.
The measurements in pA and pp($\bar{\mathrm{p}}$) collisions are 
above the model values by about a factor 4 
(corresponding to 10 experimental standard deviations; the average value 
of the measurements is 0.137$\pm$0.009, with a $\chi^2$ per degree of freedom 
of 0.88).
The relative production cross sections of charmonium states, as are
observed in all measurements in hadronic collisions cannot be described in 
the thermal approach. The temperature needed to explain the data would be
300 MeV, well above the Hagedorn limiting temperature, which is about 200 MeV.
This is in sharp contrast to the (only currently existing) measurement in 
central nucleus-nucleus collisions, performed at the SPS by the NA50 
experiment \cite{na50}, which is well described.
We recall that it was in part the observation of this measurement 
\cite{sorge97,pbm1} that brought forward the idea of statistical production
of charmed hadrons in nucleus-nucleus collisions \cite{pbm1}.
We note that the pA data exhibit a constant $\psi'/J/\psi$ production
ratio as a function of energy. 
In the model, the value is determined only by the temperature and this is 
reflected in the slight decrease of the ratio towards low energies. 
A constant value, also up to the LHC energies, is predicted
beyond $\sqrt{s_{NN}}\simeq$20 GeV.
A constant value is expected in the color evaporation model \cite{gavai}.

 The measurements reported in Fig.~\ref{fig3} demonstrate that the relative 
production cross section $\psi'$/$J/\psi$ is identical in pA and in 
pp($\bar{\mathrm{p}}$) collisions, implying no visible influence of the cold 
nuclear medium.
Note that the ratio for the Tevatron energy was derived from the CDF measurements
of $J/\psi$ \cite{cdf_jpsi} and $\psi'$ \cite{cdf_psip} and is for
transverse momentum $p_t>$1.25 GeV/c (we have extrapolated the $\psi'$ measurement
from 2 GeV/c down to 1.25 GeV/c).

\begin{figure}[htb]
\begin{tabular}{cc}
\begin{minipage}{.5\textwidth}
\hspace{-.6cm}\includegraphics[width=1.05\textwidth]{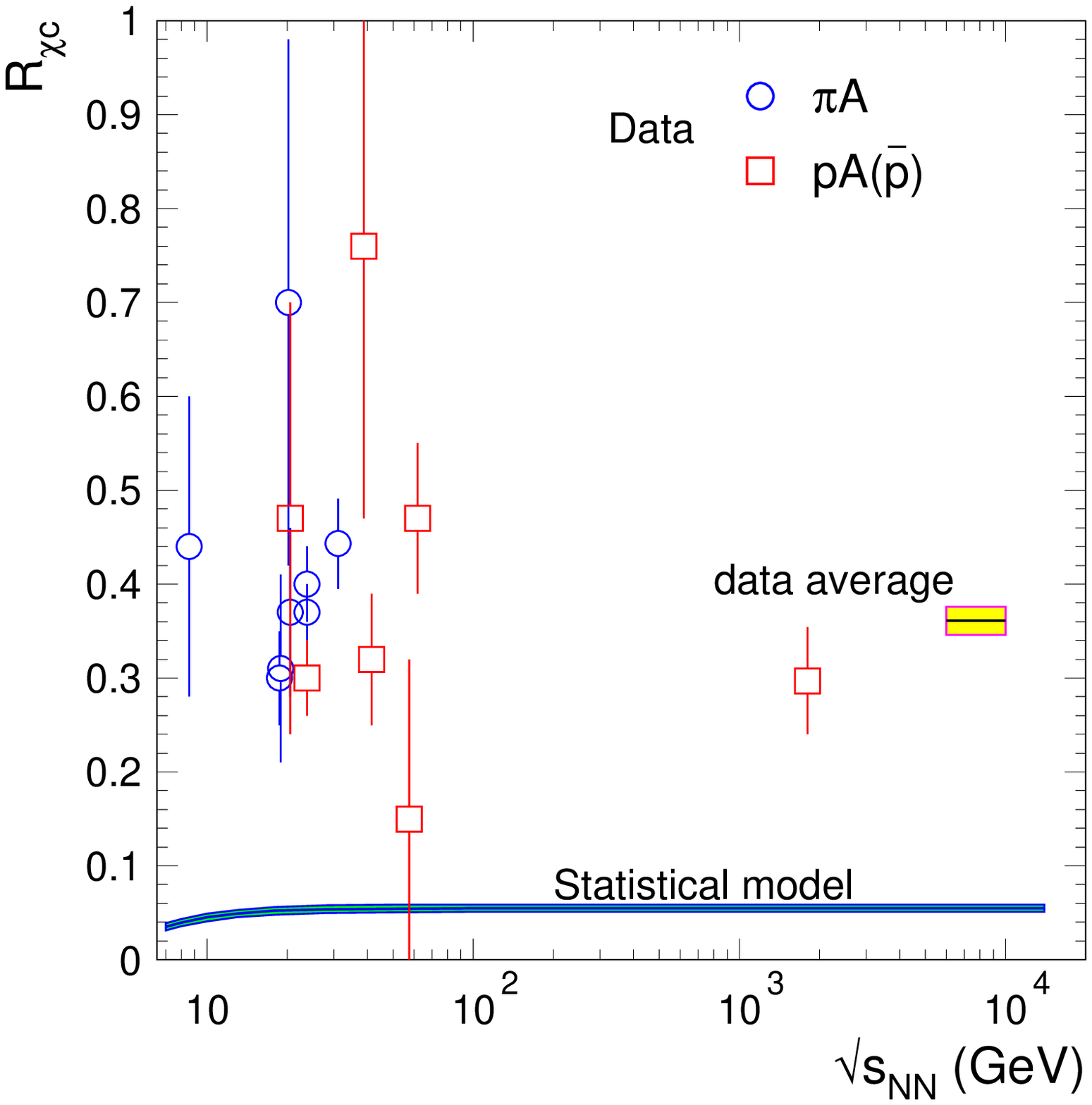}
\end{minipage} &\begin{minipage}{.5\textwidth}
\hspace{-.8cm}\includegraphics[width=1.05\textwidth]{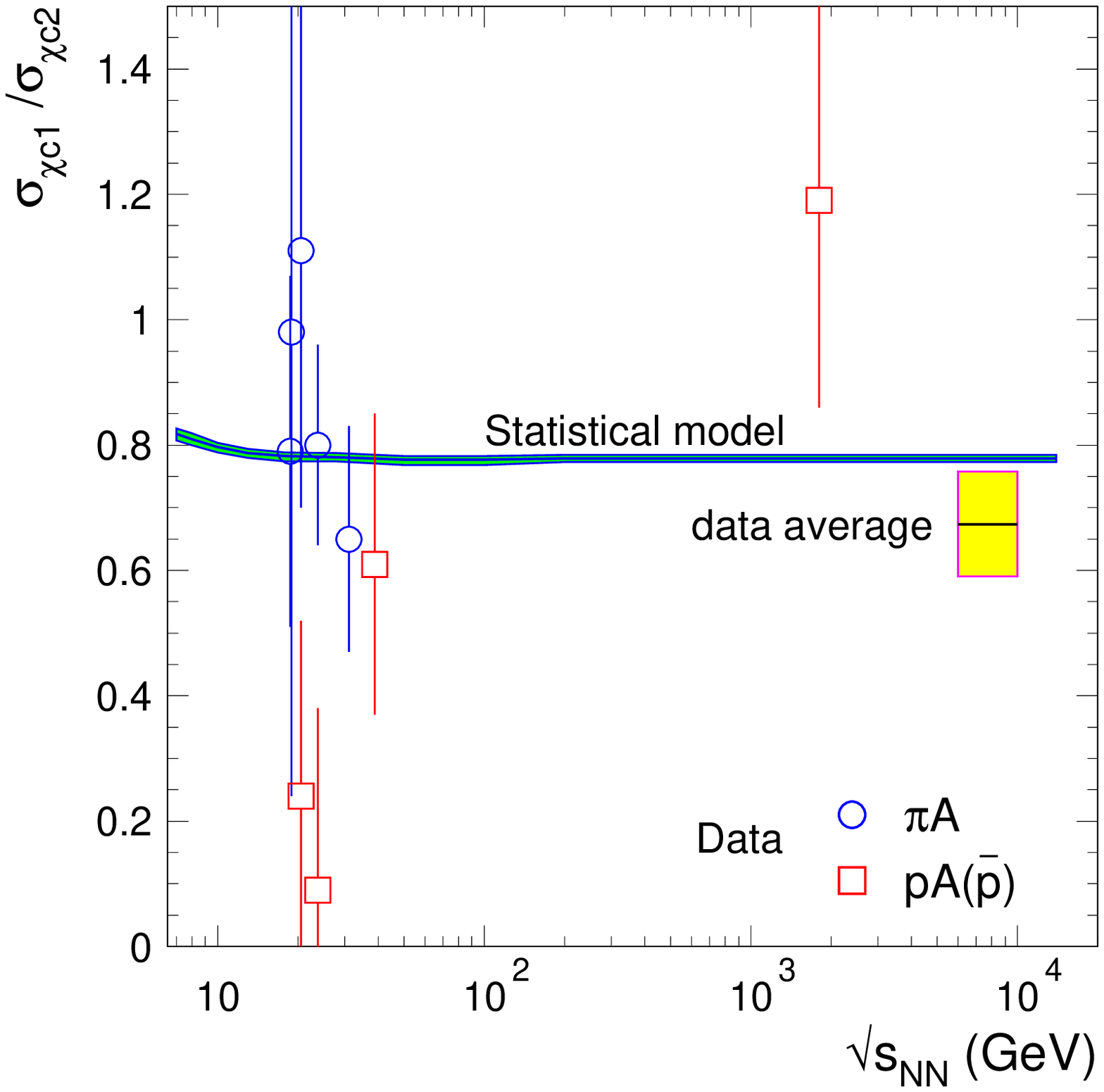}
\end{minipage}\end{tabular}
\caption{The energy dependence of the ratio $R_{\chi_c}$ (left panel) and
of the relative production cross section for $\chi_{c1}$ and $\chi_{c2}$
charmonia.
The data are form a recent compilation by the HERA-B collaboration 
\cite{herab_chic}. The average values of the measurements with the 
corresponding errors (see text) are represented by the shaded boxes.}  
\label{fig4}
\end{figure}

In Fig.~\ref{fig4} we confront the model with the data on $\chi_{c1,2}$ 
production. The data are from a recent compilation by the HERA-B collaboration 
\cite{herab_chic}. The ratio
$$R_{\chi_c} = \frac{ \sum\limits_{J=1}^2 \sigma(\chi_{cJ})Br(\chi_{cJ} \to 
J/\psi\,\gamma) }{\sigma(J/\psi)},$$
representing the fraction of $J/\psi$ mesons from radiative decays of $\chi_c$ 
states, is clearly far above the statistical model prediction.
The average value of the measurements is 0.361$\pm$0.015, with a 
$\chi^2$/dof=1.21.
An average value of 0.25$\pm$0.05 was recently determined
in an analysis of low energy data \cite{faccioli}, also well above the thermal 
value.
The relative production of the $\chi_{c1}$ and $\chi_{c2}$ charmonia, 
$\sigma_{\chi_{c1}}/\sigma_{\chi_{c2}}$, is consistent with the model, 
as can be judged from the average value of the measurements
(which is 0.674$\pm$0.084, with a $\chi^2$/dof=1.39), included in 
Fig.~\ref{fig4}.
We note that the data, characterized by rather large errors, are compatible 
as well with the expectation based on spin statistics only, which is 
$\sigma_{\chi_{c1}}/\sigma_{\chi_{c2}}$=0.6.

\begin{figure}[htb]
\centering\includegraphics[width=.53\textwidth]{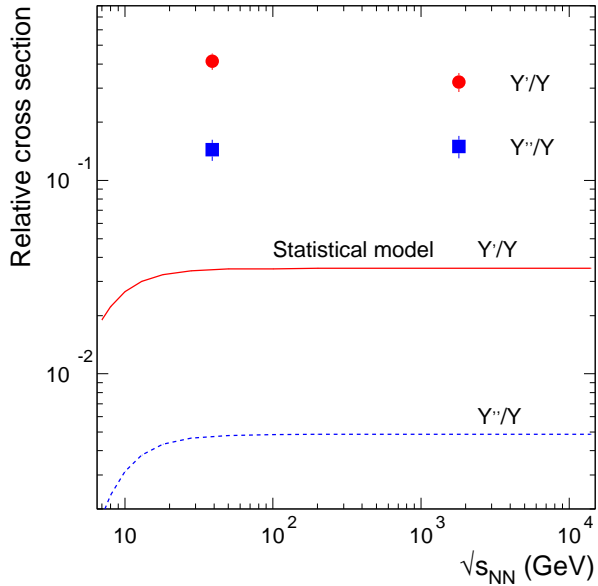}
\caption{The relative production cross section of the $\Upsilon'$ and 
$\Upsilon''$ bottomonia relative to $\Upsilon$ in pp($\bar{\mathrm p}$) collisions.
The data (symbols) are from the E866 \cite{e866_2007} and CDF \cite{cdf} 
experiments.}  
\label{fig5}
\end{figure}

We investigate the production of bottomonia in Fig.~\ref{fig5}, in terms of the
relative production cross section of the $\Upsilon'$ and $\Upsilon''$ bottomonia
relative to $\Upsilon$. The model is compared to the latest results 
for the Fermilab fixed target experiment E866 \cite{e866_2007} and in the collider 
mode from the CDF experiment \cite{cdf}. 
The data significantly exceed the thermal model calculations, by about an
order of magnitude for the $\Upsilon'/\Upsilon$ ratio 
and by almost two orders of magnitude for $\Upsilon''/\Upsilon$.
For a recent overview of the status of QCD models for hadroproduction of quarkonia
see \cite{lansb08}. 

\section{Conclusions}
We have confronted the statistical hadronization model with the most recent 
data on the production of open heavy flavor hadrons and quarkonia in e$^+$e$^-$ 
and in pp and p($\pi$)-nucleus collisions. 
Employing the parameters extracted from the analyses of light flavor hadron 
production the model describes well the fragmentation to open heavy flavor hadrons.
In contrast, quarkonium production cannot be described in this framework.
We emphasize again that, for the e$^+$e$^-$ collisions, we have employed
a canonical treatment of the two $q$ and $\bar{q}$ jets. The relative
abundance of the five flavors, taken from the measurements at the $Z^0$ 
resonance, are external input values unrelated with the thermal model.
We note that, in contrast to the hadrons carrying strangeness, for which
a strangeness suppression factor $\gamma_s<1$ is the outcome of the fit 
for e$^+$e$^-$ as well as for pp collisions, charmed and bottom hadrons 
are described in e$^+$e$^-$ collisions without any other parameter.
The strange quark, with its intermediate mass, which is comparable to $T$, 
seems to have an interesting intermediate status in between the very light 
and the very heavy quarks.
In the hadronization process, the $u$ and $d$ quarks reach abundances
consistent with a thermal ensemble for $T\simeq$165 MeV. 
On the other side, the $c$ and $b$ quark production is determined by either 
pQCD (in pp) or electroweak (in e$^+$e$^-$) processes and the thermal
model only describes hadronization phase space.
In contrast, there is a significant number of newly produced $s$ and $\bar{s}$
quarks, but incomplete equilibration leads to $\gamma_s<1$ implying that $s$
quark abundances are not thermal.
Remarkably, strangeness suppression is lifted for central nucleus-nucleus 
collisions, implying full equilibration.

The fact that, in elementary collisions, quarkonium production cannot be described
by the statistical model is in sharp contrast to the situation in 
nucleus-nucleus collisions, where all the measurements to date are well 
described by statistical hadronization. 
In general, statistical production can only take place effectively if the 
charm quarks reach thermal equilibrium and are free to travel over a large 
distance, implying deconfinement. 
The model will be most dramatically tested at the LHC energies,
where data are expected within a year.
If confirmed, statistical production of charmed (and possibly also bottom)
hadrons, in particular of J/$\psi$, will provide a crucial determination
of the QCD phase boundary.
Whether the bottom quarks equilibrate in the QGP is an important open question,
which will be addressed by the measurements at the LHC.
Predictions exist, both within the statistical approach \cite{aa03,kuz} 
as well as within a kinetic approach \cite{grand}.

Acknowledgements:  
We acknowledge the support of the Alliance Program of the Helmholtz 
Association HA216/EMMI.
K.R. acknowledges partial support from the Polish Ministry of Science (MENiSW)
and the Deutsche Forschungsgemeinschaft (DFG) under the Mercator Programme.

\end{document}